\documentclass[twoside,a4paper,11pt]{sea9}
\usepackage{graphicx}
\usepackage{hyperref}
\usepackage{movie15}
\begin{document}
\pagenumbering{arabic}
\pagestyle{myheadings}
\thispagestyle{empty}
{\flushleft\includegraphics[width=\textwidth,bb=58 650 590 680]{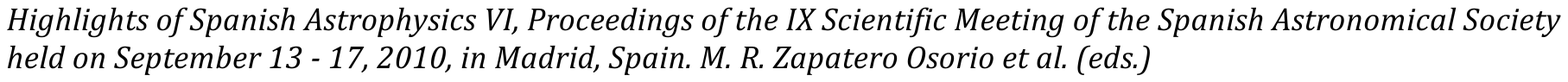}}
\vspace*{0.2cm}
\begin{flushleft}
{\bf {\LARGE
%
The pipeline for the GOSSS data reduction
%
}\\
\vspace*{1cm}
%
Alfredo Sota$^{1}$, 
and
Jes\'us Ma\'{\i}z Apell\'aniz$^{1}$
%
}\\
\vspace*{0.5cm}
%
$^{1}$
Instituto de Astrof\'{\i}sica de Andaluc\'{\i}a-CSIC, Glorieta de la Astronom\'{\i}a s/n, 18008 Granada, Spain 
%
\end{flushleft}
%
\markboth{
The pipeline for the GOSSS data reduction
}{ 
%
Alfredo Sota \& Jes\'us Ma\'{\i}z Apell\'aniz
%
}
\thispagestyle{empty}
\vspace*{0.4cm}
\begin{minipage}[l]{0.09\textwidth}
\ 
\end{minipage}
\begin{minipage}[r]{0.9\textwidth}
\vspace{1cm}
\section*{Abstract}{\small
%
The Galactic O-Star Spectroscopic Survey (GOSSS) is an ambitious project that is observing all known Galactic O stars with $B$ $<$ 13 in 
the blue-violet part of the spectrum with R$\sim$2500.  It is based on version 2 of the most complete catalog to date of Galactic O stars 
with accurate spectral types (v1, Ma\'{\i}z Apell\'aniz et al. 2004 ;v2, Sota et al. 2008). Given the large amount of data that we are 
getting (more than 150 nights of observations at three different observatories in the last 4 years) we have developed an automatic spectroscopic 
reduction pipeline. This pipeline has been programmed in IDL and automates the process of data reduction. It can operate in two modes: 
automatic data reduction (quicklook) or semi-automatic data reduction (full). In "quicklook", we are able to get rectified and calibrated 
spectra of all stars of a full night just minutes after the observations. The pipeline automatically identifies the type of image and applies 
the standard reduction procedure (bias subtraction, flat field correction, application of bad pixel mask, ...). It also extracts all spectra of the 
stars in one image (including close visual binaries), aligns and merges all spectra of the same star (to increase the signal to noise ratio and 
to correct defects such as cosmic rays), calibrates in wavelength and rectifies the continuum. The same operations are performed in full mode, 
but allowing the user to adjust the parameters used in the process.
%
\normalsize}
\end{minipage}
\end{document}